\documentclass{appolb}
\usepackage{graphicx}

\usepackage{multirow}
\usepackage{ctable}
\usepackage{booktabs}
\usepackage{array}
\usepackage{tabularx}
\usepackage{xcolor}
\usepackage{pstricks}

\def\Pom{{\bf I\!P}}
\def\Reg{{\bf I\!R}}

\begin{document}
\title{Single- and central-diffractive production\\ of open charm and bottom mesons at the LHC%
\thanks{Presented at the 16th conference on Elastic and Diffractive scattering, EDS Blois 2015 }%
}
\author{Marta {\L}uszczak, Rafa{\l} Maciu{\l}a and Antoni Szczurek
\address{University of Rzesz\'ow, Rejtana 16, 35-959 Rzesz\'ow, Poland \and
H.Niewodnicza\'nski Institute of Nuclear Physics, Polish Academy of Sciences, Radzikowskiego 152, 31-342 Krak\'ow, Poland}
}
\maketitle
\begin{abstract}
We discuss diffractive production of open charm and bottom mesons at the LHC.
The differential cross sections for single- and central-diffractive mechanisms for $c\bar c$ and $b\bar b$ pair production are calculated in the framework of the Ingelman-Schlein model corrected for absorption effects. 
The LO gluon-gluon fusion and quark-antiquark
anihilation partonic subprocesses are taken into consideration, which are calculated within
standard collinear approximation. The extra corrections from reggeon exchanges are
taken into account. 
The hadronization of charm and bottom quarks is taken into account by means of fragmentation functions.
Predictions for single- and central-diffractive production in the case of $D$ and $B$ mesons,
as well as $D\bar D$ pairs are presented, including detector acceptance of the ATLAS, CMS and LHCb Collaborations.
\end{abstract}

\section{Introduction}
On theoretical side diffractive processes are related 
with exchange of pomeron or processes with the QCD amplitude without net
color exchange. In such processes pomeron must be treated rather technically,
depending on the formulation of the approach.
Experimentally such processes are defined by special requirement(s)
on the final state. The most popular is a requirement of rapidity gap
starting from the final proton(s) on one (single-diffrative process) or both
(central-diffractive process) sides. 
Several processes with different final states were studied
at HERA, such as dijet, charm production, etc. 
The H1 Collaboration has found a set of so-called
diffractive parton distributions in the proton inspired by 
the Ingelman-Schlein model \cite{IS}, which we will use in the presented
studies. In this fit both pomeron 
and reggeon contributions were included.

In hadronic processes so far only some selected diffractive processes
were discussed in 
the literature such as diffractive production of dijets \cite{Kramer}, 
production of $W$ \cite{Collins} and $Z$ \cite{CSS09} bosons, 
production of $W^+ W^-$ pairs \cite{LSR} or production of 
$c \bar c$ \cite{LMS2011}. 
The latter was done there only for illustration of the general situation
at the parton level.
The cross section for diffractive processes are in general rather
small, (e.g. the single-diffractive processes are of the order of 
a few percent compared to inclusive cross sections).

\section{Theoretical framework}

\begin{figure}[!ht]
\begin{center}
\begin{flushleft}
{\scriptsize a) \hskip+77mm b)}
\end{flushleft} \vskip-5mm
\includegraphics[height=.11\textheight]{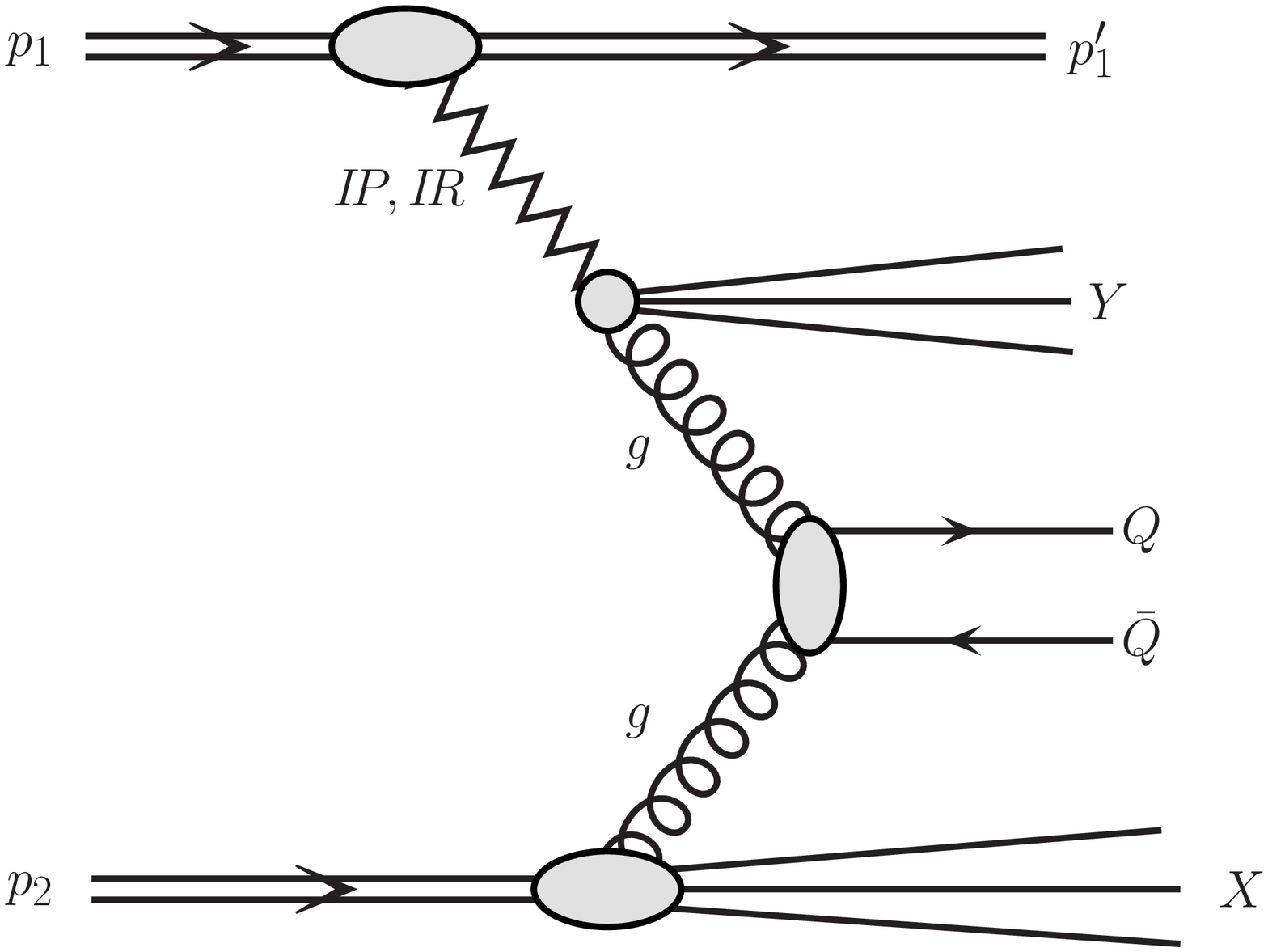}\hskip-1mm
\includegraphics[height=.11\textheight]{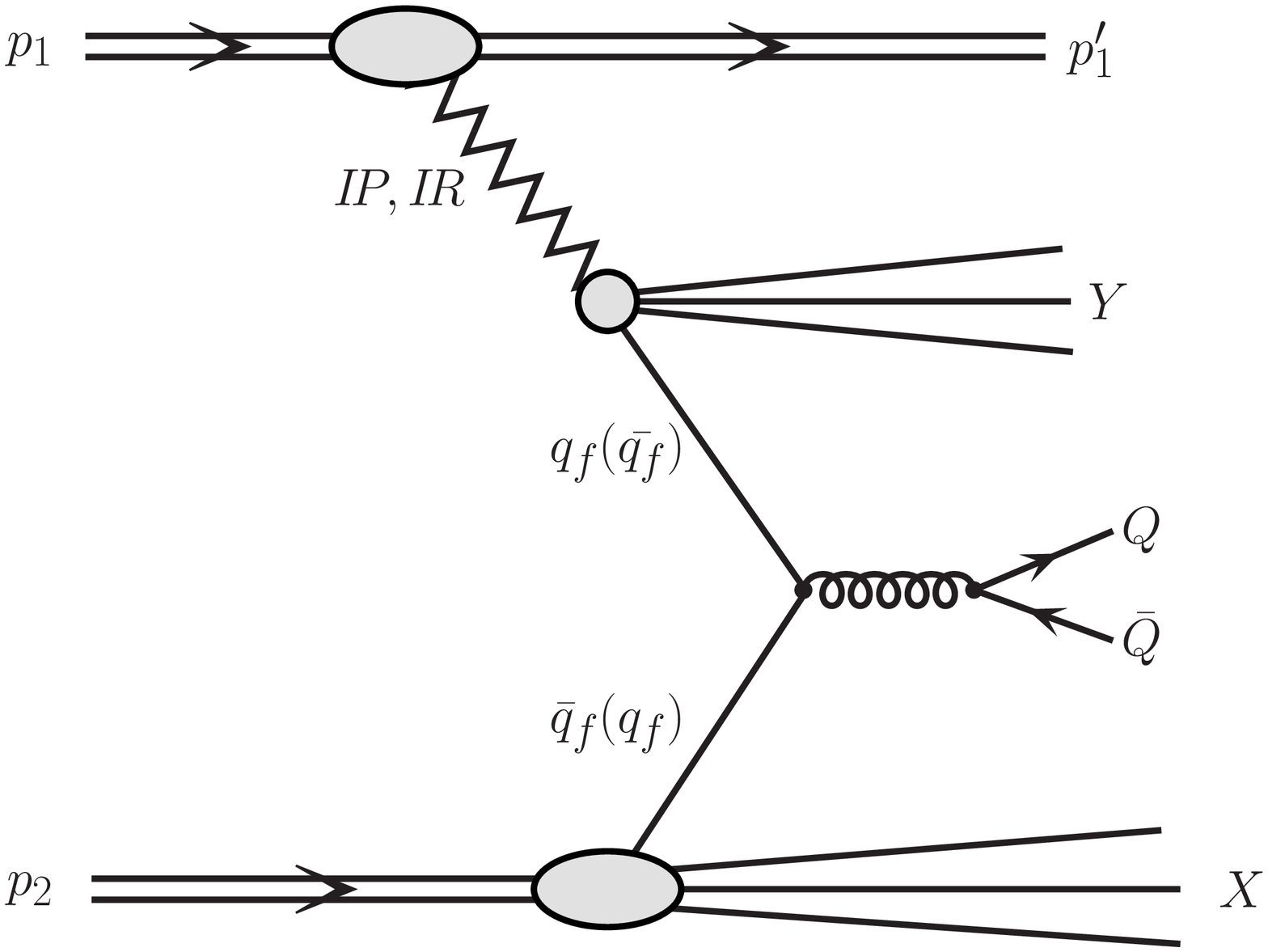}\hskip-1mm
\includegraphics[height=.11\textheight]{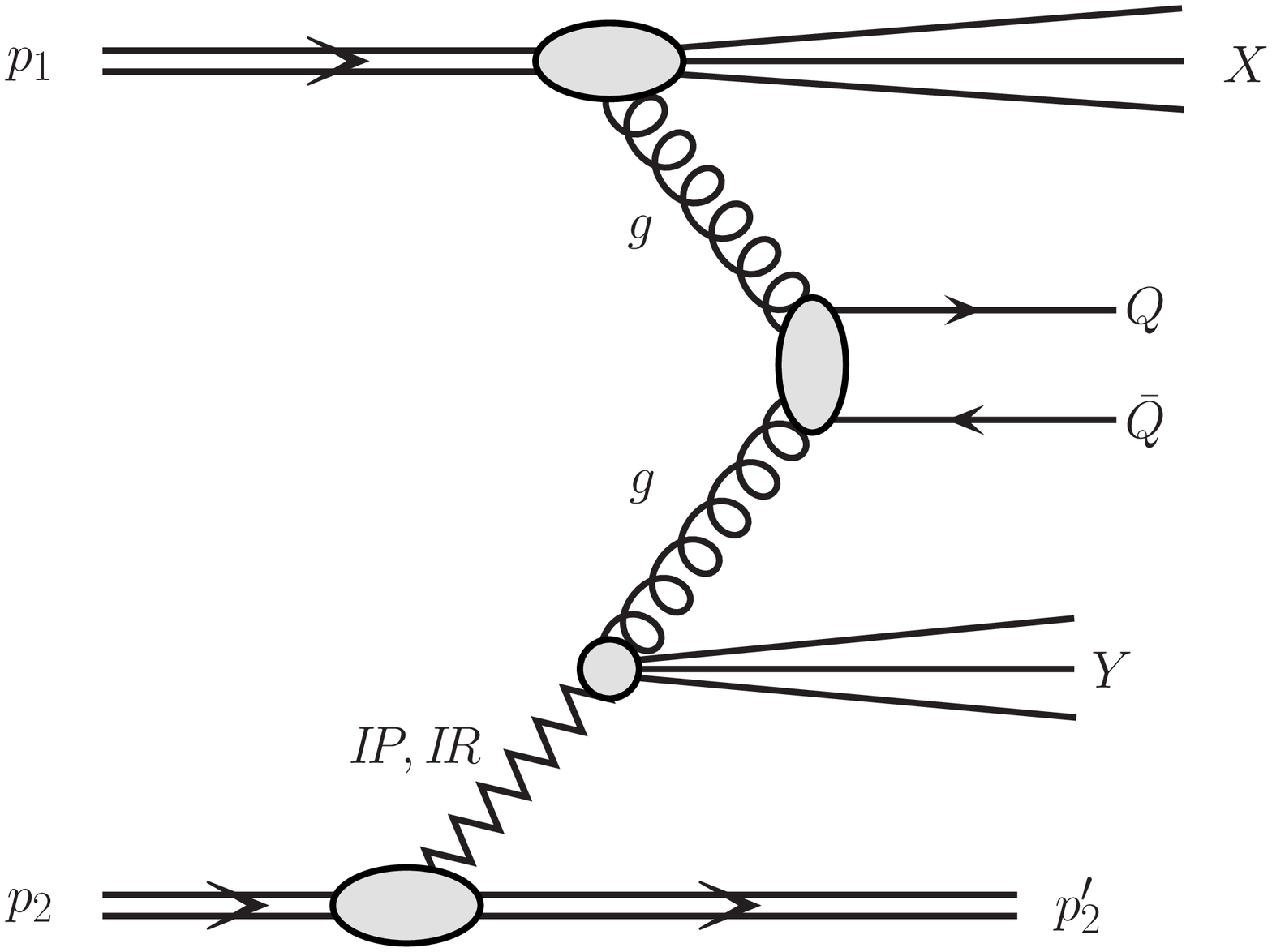}\hskip-1mm
\includegraphics[height=.11\textheight]{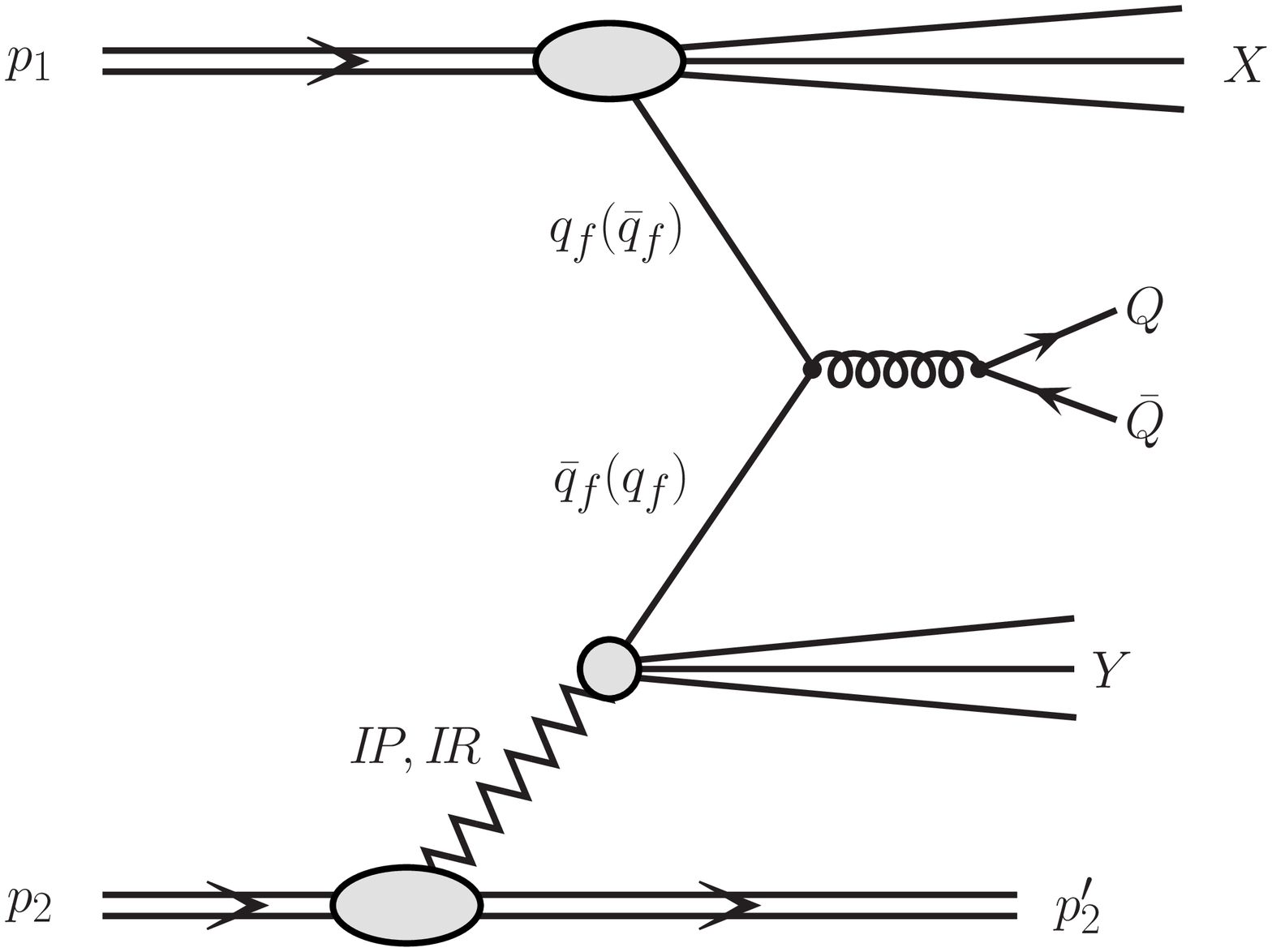}
\end{center}
\caption{
The mechanisms of single-diffractive production of heavy quarks.
}
 \label{fig:1}
\end{figure}

\begin{figure}[!ht]
\begin{center}
\includegraphics[height=.13\textheight]{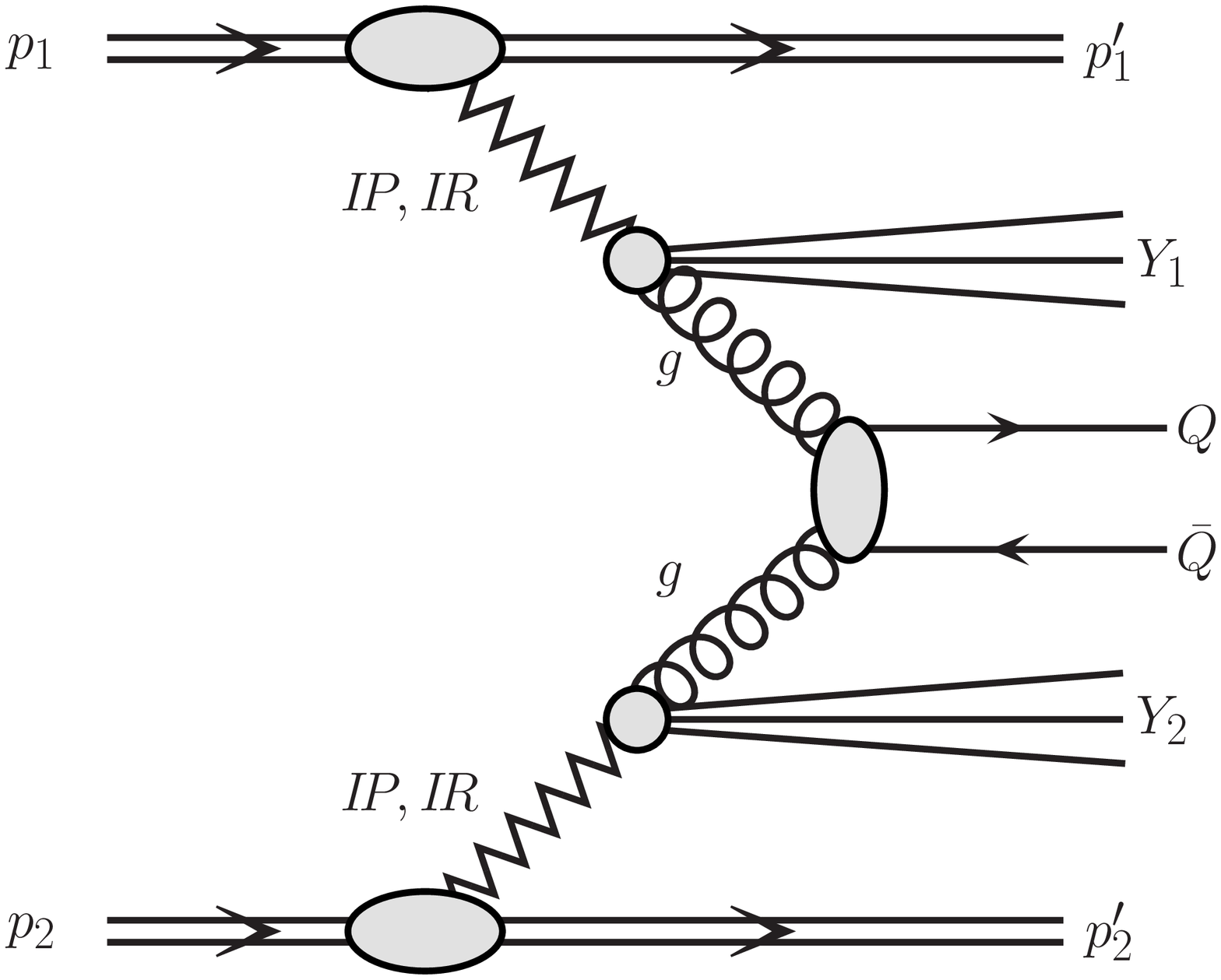}
\includegraphics[height=.13\textheight]{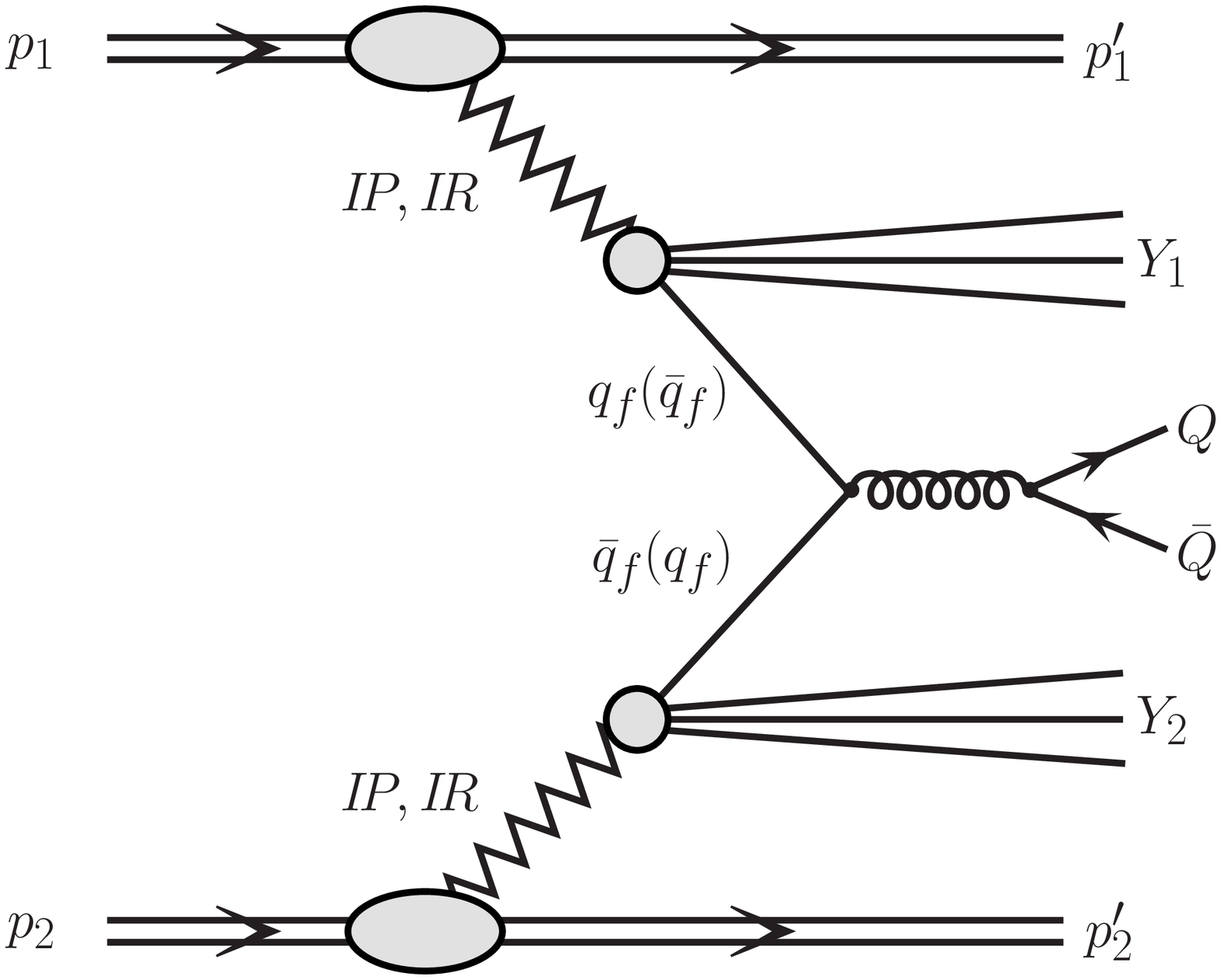}
\end{center}
\caption{
The mechanisms of central-diffractive production of heavy quarks. 
}
\label{fig:2}
\end{figure}

The mechanisms of the diffractive production 
of heavy quarks ($c \bar c$, $b \bar b$) discussed here are shown in 
Figs.~\ref{fig:1} and ~\ref{fig:2}. Both, LO
gg-fusion and $q \bar q$-anihilation partonic subprocesses
are taken into account in the calculations.

In the following we apply the Ingelman-Schlein approach \cite{IS}. 
Details of our calculations of corresponding differential cross sections
can be found in Ref. \cite{LMS2015}.

\subsection{Results for diffractive heavy quarks pair production}

\begin{figure}[!h]
\begin{minipage}{0.47\textwidth}
 \centerline{\includegraphics[width=1.0\textwidth]{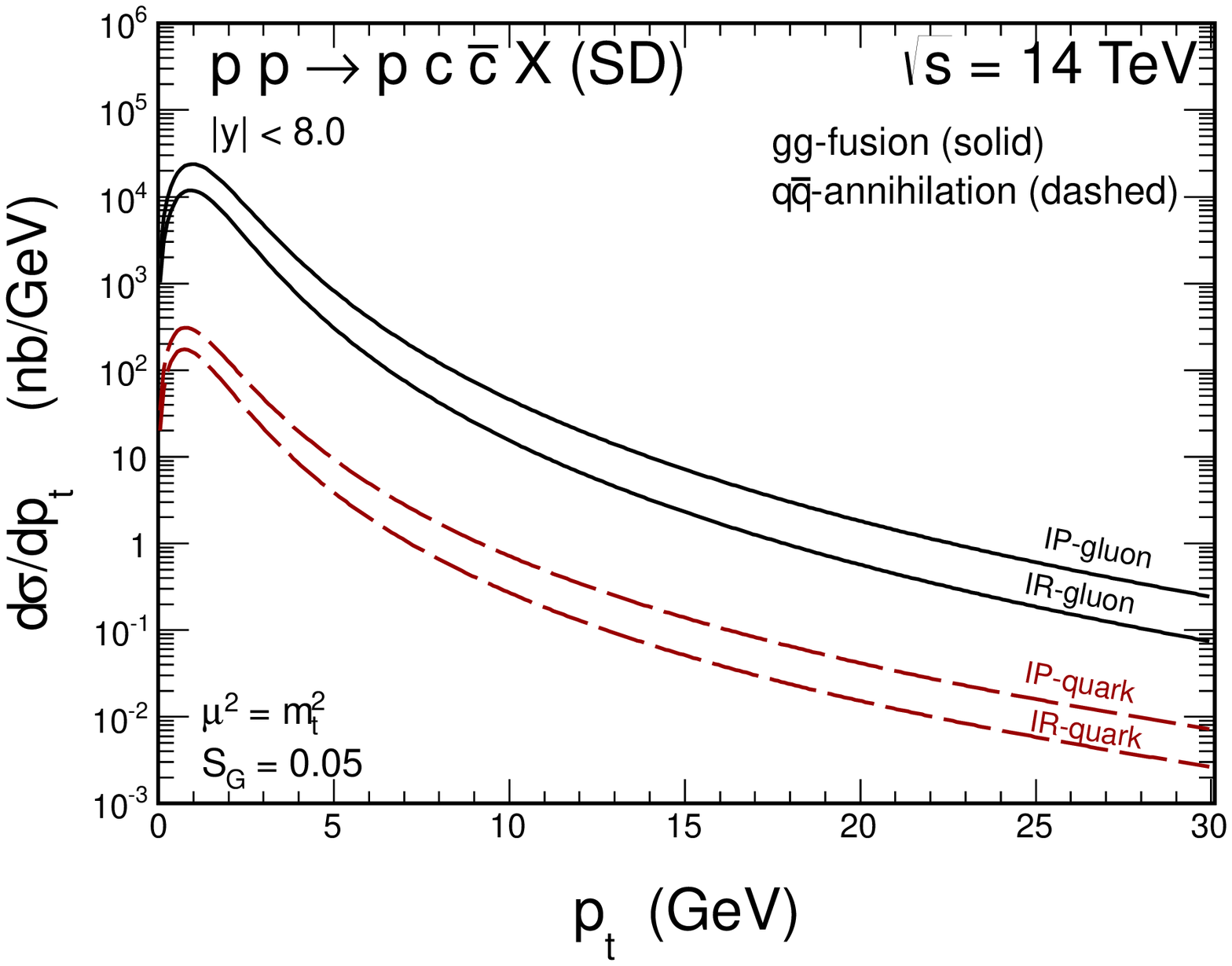}}
\end{minipage}
\hspace{0.5cm}
\begin{minipage}{0.47\textwidth}
 \centerline{\includegraphics[width=1.0\textwidth]{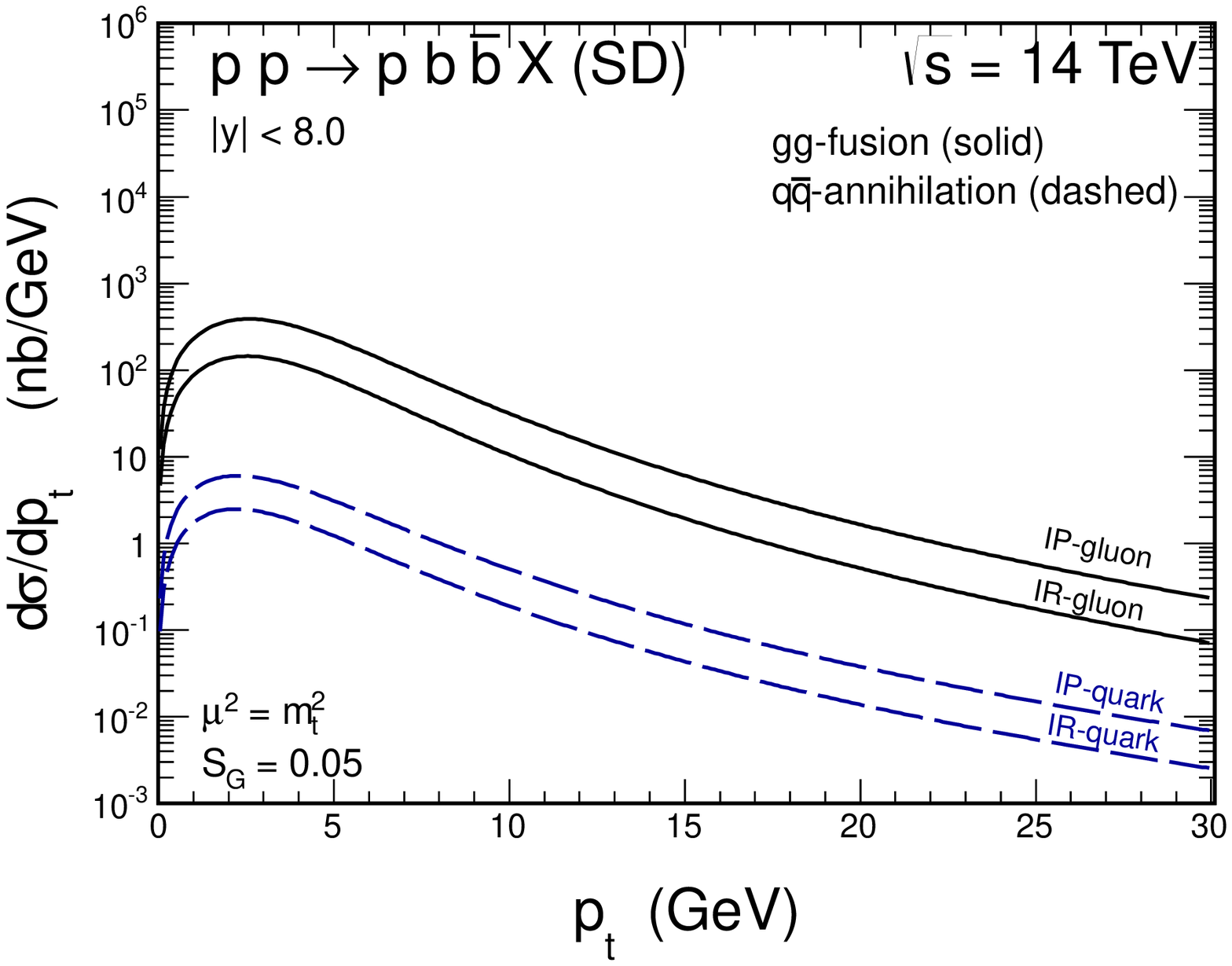}}
\end{minipage}
   \caption{
\small Transverse momentum distribution of $c$ quarks (antiquarks) (left)
and $b$ quarks (antiquarks) (right) for single-diffractive production at $\sqrt{s} = 14$ TeV.
}
 \label{fig:pt}
\end{figure}

In Fig.~\ref{fig:pt} we show the transverse momentum distribution of $c$ quarks (antiquarks)
and $b$ quarks (antiquarks) for single-diffractive production at $\sqrt{s} = 14$ TeV. 
Components of the pomeron-gluon (and gluon-pomeron)
are almost two orders of magnitude larger than
the pomeron-quark(antiquark) and quark(antiquark)-pomeron.
The estimated reggeon contribution is slightly smaller.

Different models of absorption corrections 
(one-, two- or three-channel approaches) 
for diffractive processes were presented in the literature.
The absorption effects for the diffractive processes were calculated e.g.
in \cite{KMR2000,Maor2009,CSS09}.
The different models give slightly different predictions.
Usually an average value of the gap survival probability
$<|S_G|^2>$ is calculated first and then the cross sections for different
processes is multiplied by this value.
We follow this somewhat simplified approach.
Numerical values of the gap survival probability can be found 
in \cite{KMR2000,Maor2009,CSS09}.
The multiplicative factors are $S_G$ = 0.05
for single-diffractive production 
and $S_G$ = 0.02 for central-diffractive one for the nominal LHC energy ($\sqrt{s}=$ 14 TeV).

\begin{figure}[!h]
\begin{minipage}{0.47\textwidth}
 \centerline{\includegraphics[width=1.0\textwidth]{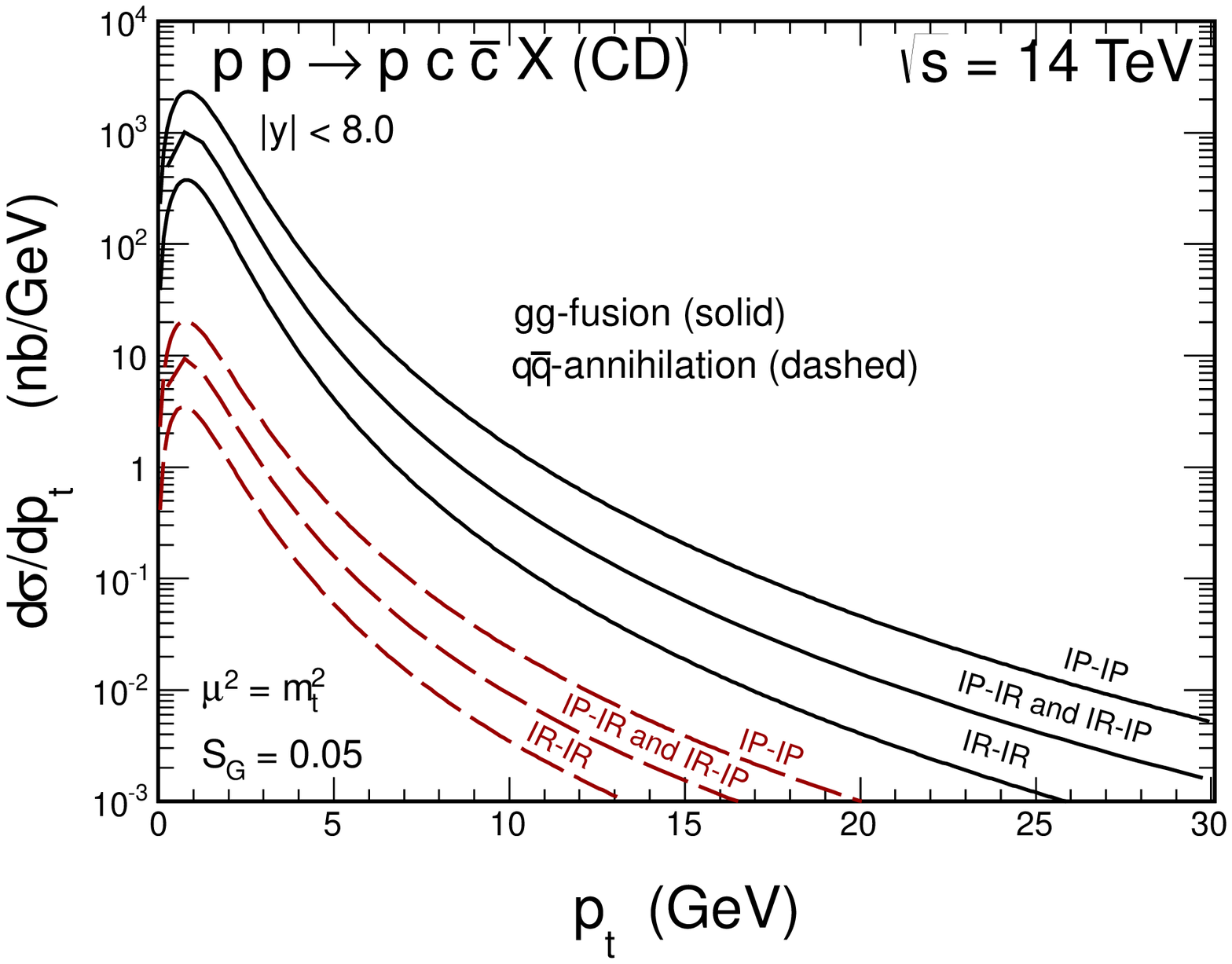}}
\end{minipage}
\hspace{0.5cm}
\begin{minipage}{0.47\textwidth}
 \centerline{\includegraphics[width=1.0\textwidth]{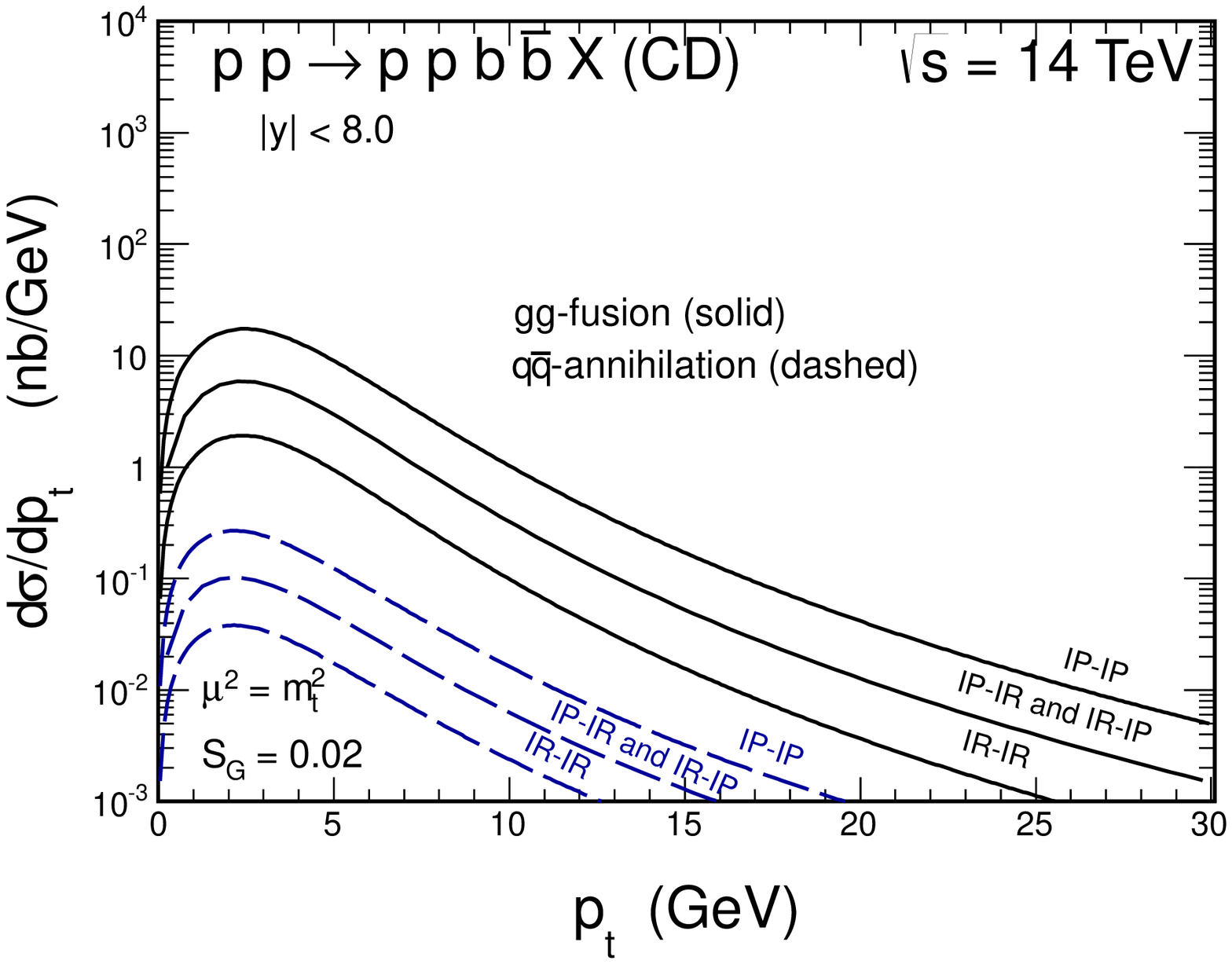}}
\end{minipage}
   \caption{
\small Transverse momentum distribution of $c$ quarks (antiquarks) (left)
and $b$ quarks (antiquarks) (right) for the central-diffractive production at $\sqrt{s} = 14$ TeV.
}
 \label{fig:pt_CD}
\end{figure}

In Fig.~\ref{fig:pt_CD} we show the transverse momentum distribution of
$c$ quarks (antiquarks)
and $b$ quarks (antiquarks) for central-diffractive production at $\sqrt{s} = 14$ TeV.
The distributions for central-dffractive component is smaller than that for the single-diffractive distributions by almost two orders of magnitude. 

\begin{figure}[!h]
\begin{minipage}{0.47\textwidth}
 \centerline{\includegraphics[width=1.0\textwidth]{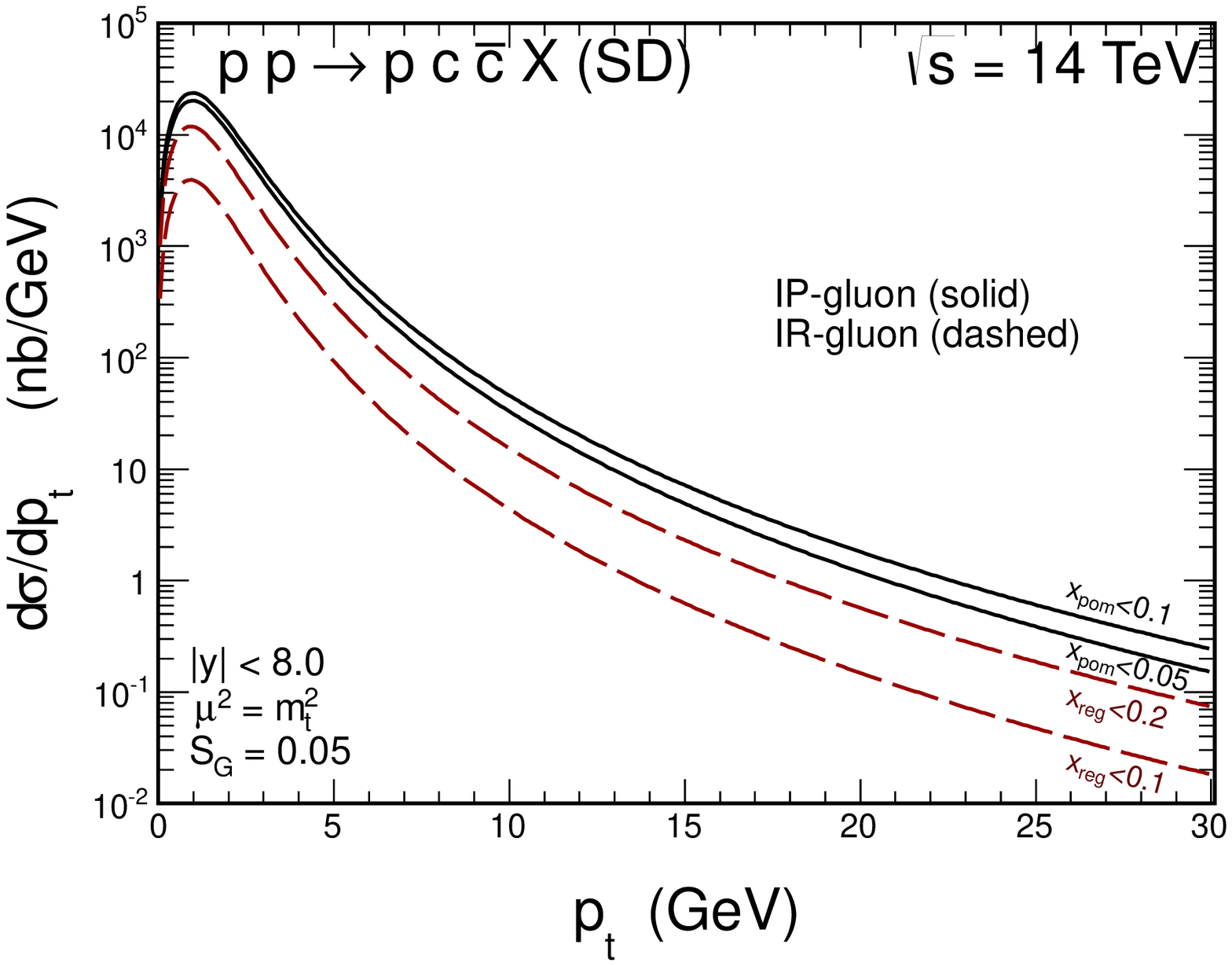}}
\end{minipage}
\hspace{0.5cm}
\begin{minipage}{0.47\textwidth}
 \centerline{\includegraphics[width=1.0\textwidth]{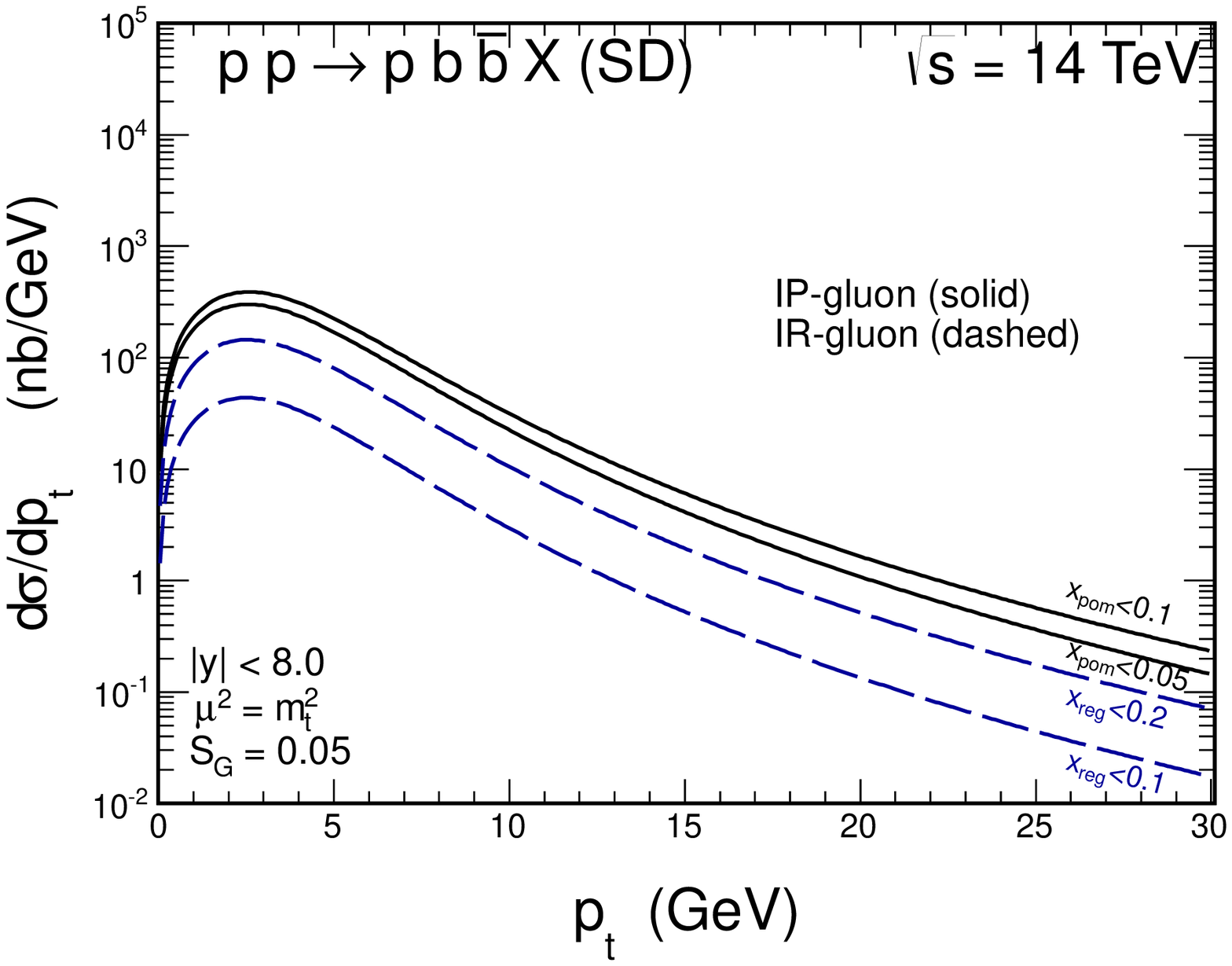}}
\end{minipage}
   \caption{
\small Transverse momentum distribution of $c$ quarks (antiquarks) (left)
and $b$ quarks (antiquarks) (right) for single-diffractive production at $\sqrt{s} = 14$ TeV for different maximal $x_\Pom$ (solid) and $x_\Reg$ (dashed). 
}
 \label{fig:pt_x}
\end{figure}

In Fig.~\ref{fig:pt_x} we show separately contributions for different upper limits for the value of $x_\Pom$ and $x_\Reg$. The shape of these distributions are rather similar.
As a default, in the case of pomeron exchange the upper limit in 
the convolution formula is taken to be 0.1 and for reggeon exchange 0.2.

\begin{figure}[!h]
\begin{minipage}{0.47\textwidth}
 \centerline{\includegraphics[width=1.0\textwidth]{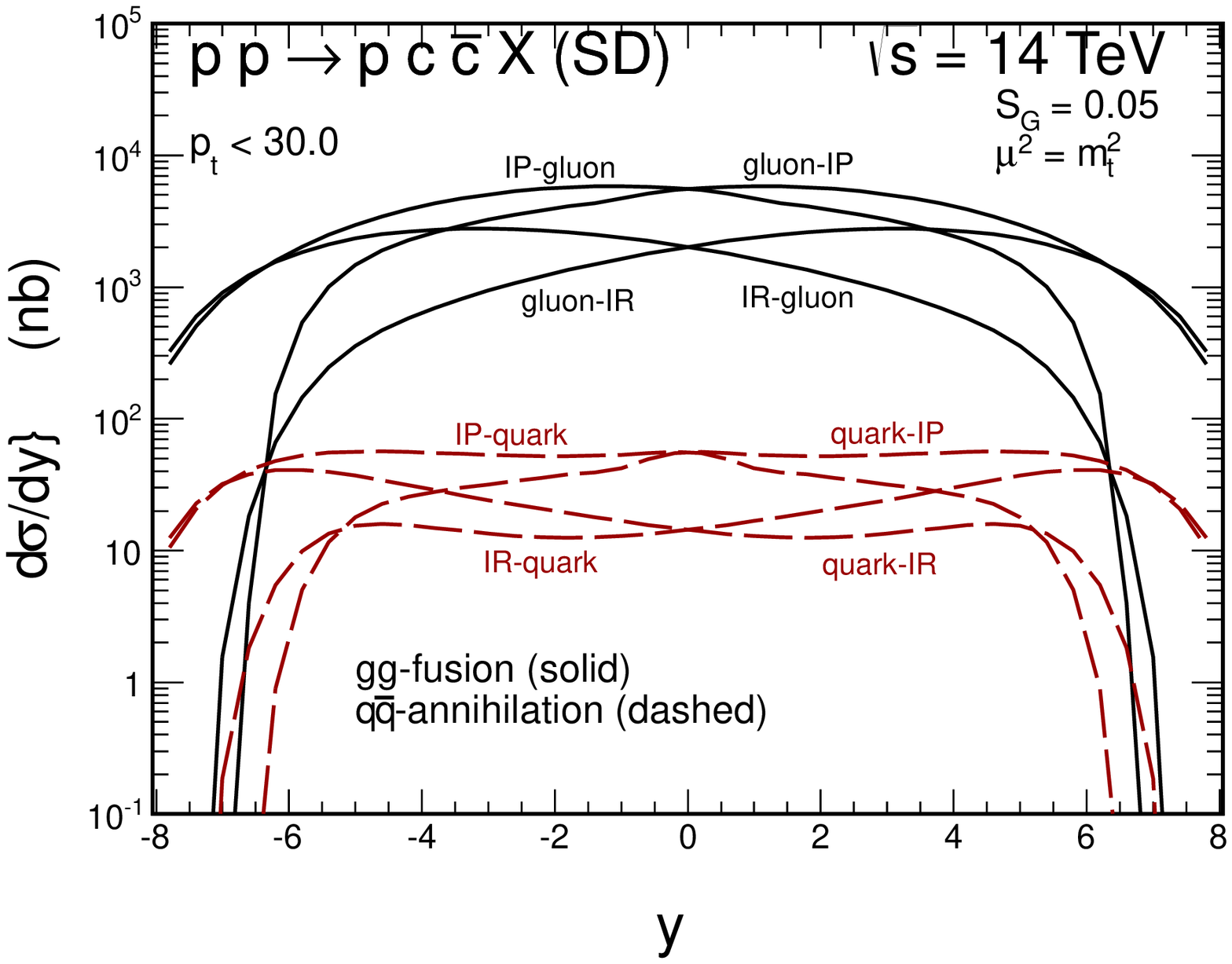}}
\end{minipage}
\hspace{0.5cm}
\begin{minipage}{0.47\textwidth}
 \centerline{\includegraphics[width=1.0\textwidth]{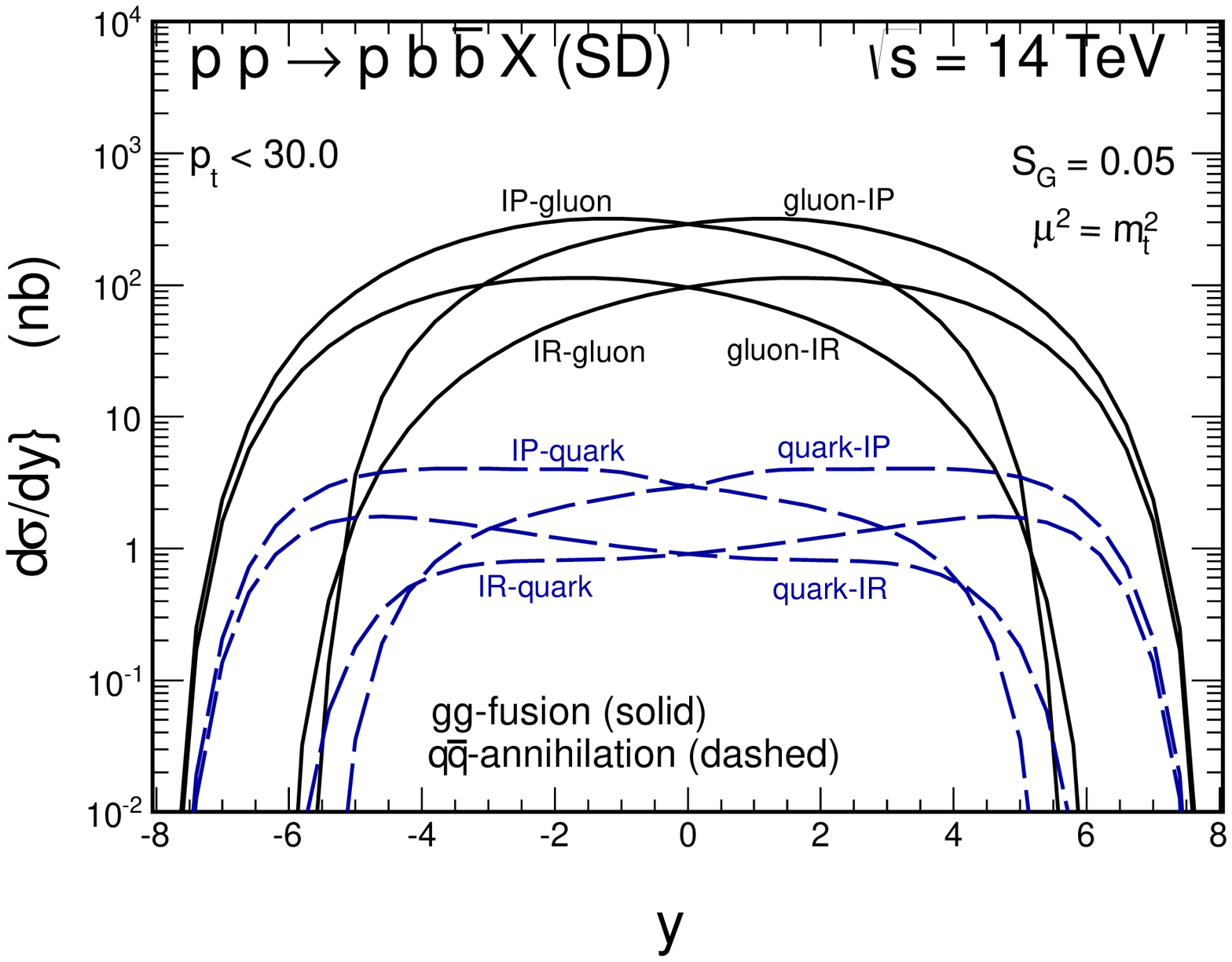}}
\end{minipage}
   \caption{
\small Rapidity distribution of $c$ quarks (antiquarks) (left)
and $b$ quarks (antiquarks) (right) for single-diffractive production at $\sqrt{s} = 14$ TeV.
}
 \label{fig:y}
\end{figure}

\begin{figure}[!h]
\begin{minipage}{0.47\textwidth}
 \centerline{\includegraphics[width=1.0\textwidth]{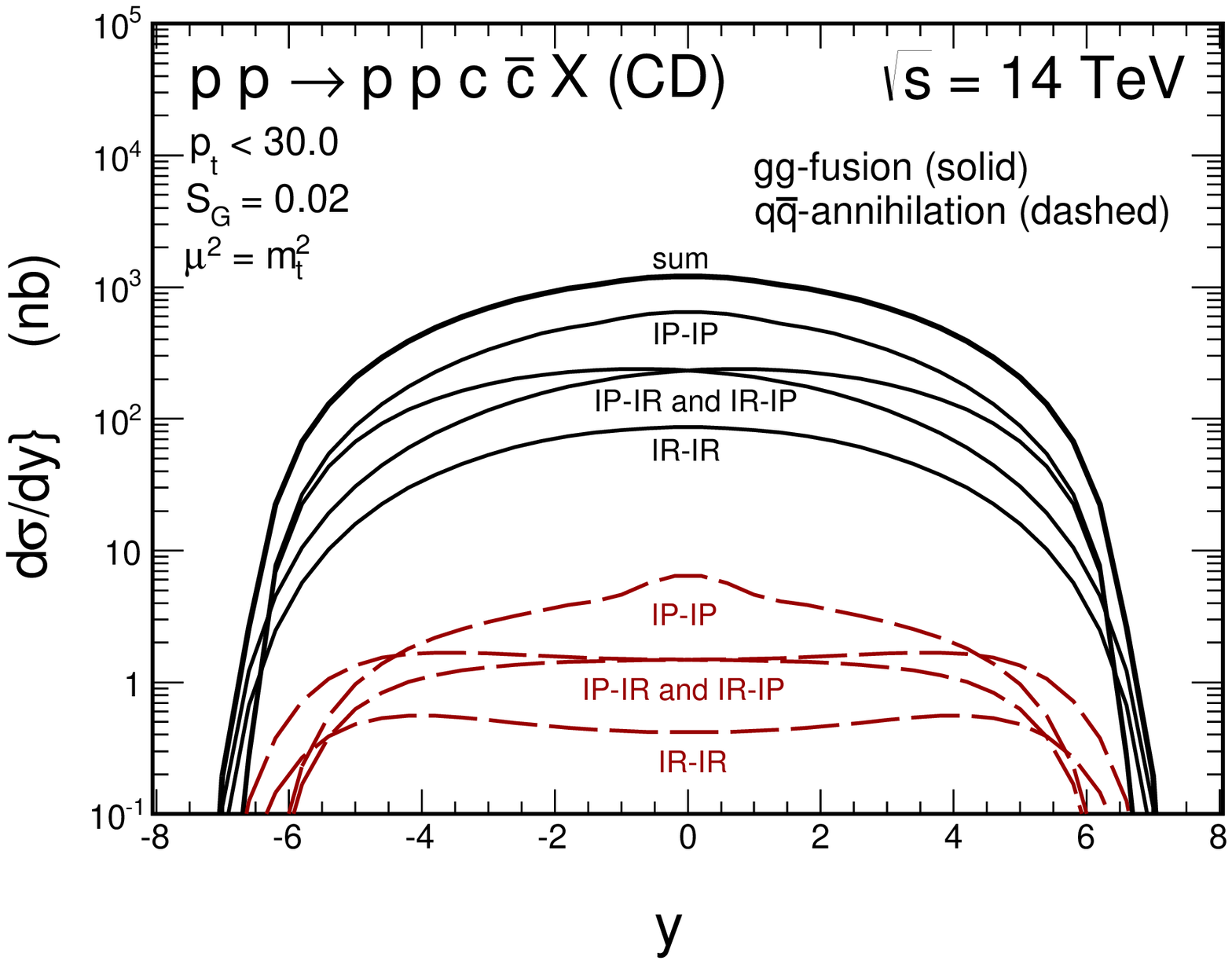}}
\end{minipage}
\hspace{0.5cm}
\begin{minipage}{0.47\textwidth}
 \centerline{\includegraphics[width=1.0\textwidth]{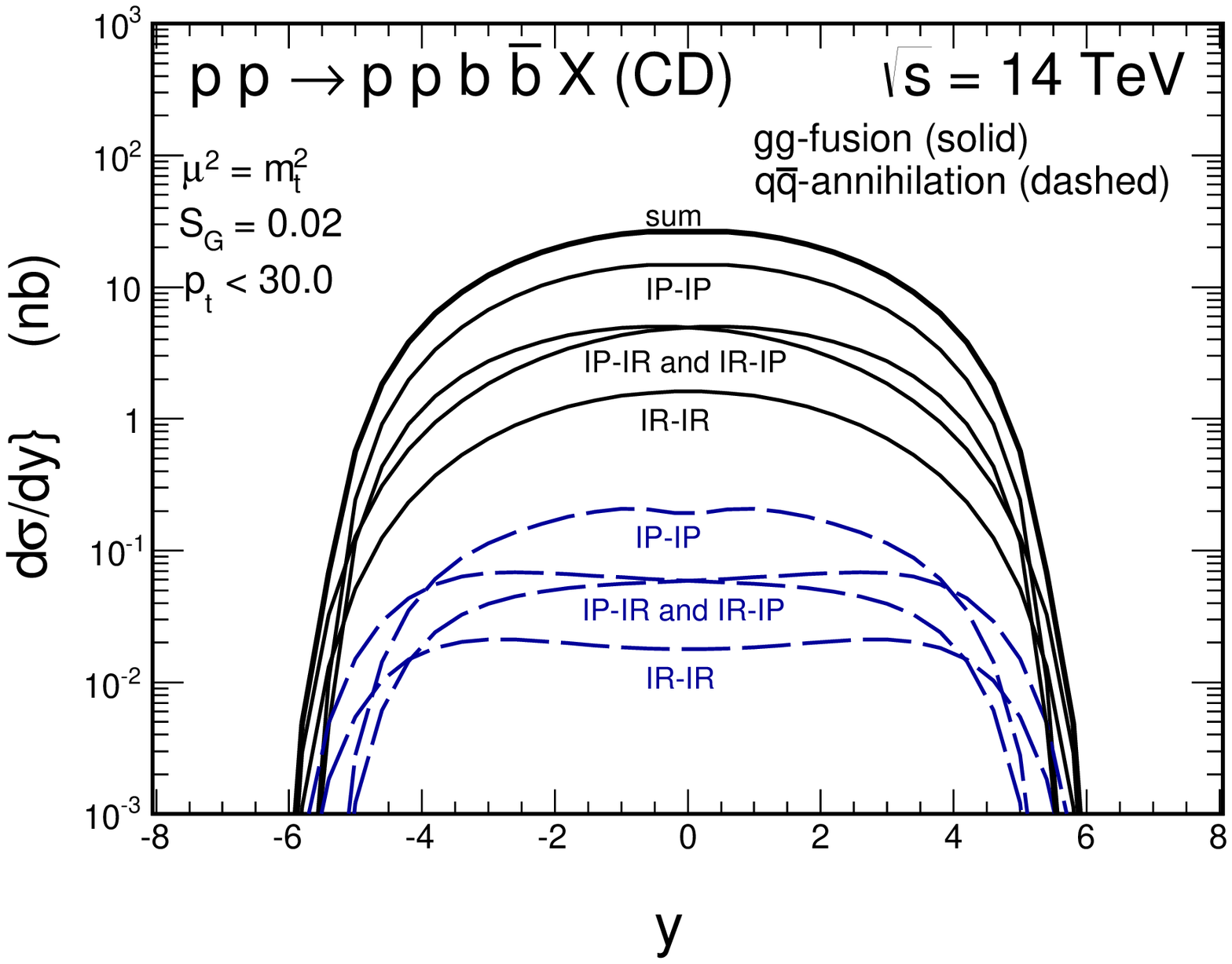}}
\end{minipage}
   \caption{
\small Rapidity distribution of $c$ quarks (antiquarks) (left)
and $b$ quarks (antiquarks) (right) for the central-diffractive production at $\sqrt{s} = 14$ TeV.
}
 \label{fig:y_CD}
\end{figure}

Figures~\ref{fig:y} and~\ref{fig:y_CD} show rapidity distributions for
$c$ quarks (antiquarks) (left panels)
and $b$ quarks (antiquarks) (right panels) production for single- and central-diffractive mechanisms, respectively. 
The rapidity distributions for pomeron-gluon (and gluon-pomeron), pomeron-quark(antiquark) (and quark(antiquark)-pomeron) and reggeon-gluon (and gluon-reggeon),
reggeon-quark(antiquark) (and quark(antiquark)-reggeon)
mechanisms in the single-diffractive case are shifted
to forward and backward rapidities, respectively. 
The distributions for the individual single-diffractive
mechanisms have maxima at large rapidities, while
the central-diffractive contribution is concentrated
at midrapidities. This is a consequence of limiting integration:
0.0 $< x_\Pom <$ 0.1 
and $x_\Reg$ to 0.0 $< x_\Reg <$ 0.2.

\subsection{Cross sections for $D^{0}$ and $B^{\pm}$ mesons production}

Measurements of charm and bottom cross sections at hadron colliders is based on full reconstruction of all decay products of open charm and bottom mesons, for instance in the 
$D^0 \to K^- \pi^+ $, $D^+ \to K^- \pi^+ \pi^+$ or $B^+ \to J/\psi K^+ \to K^+ \mu^+ \mu^-$ channels. The decay products
with an invariant mass from the expected hadron decay combinations, permit direct observation of $D$ or $B$ meson as a peak in relevant invariant mass spectrum. Then, after a substraction of invariant mass continuum background the relevant cross section for the meson production is obtained.
The same method can be applied for measurement of charm and bottom production rates for the diffractive events.

\begin{table}[tb]%
{\large
\caption{Integrated cross sections for diffractive production of open charm and bottom mesons in different measurement modes for ATLAS, LHCb and CMS experiments at $\sqrt{s}=14$ TeV.}
\newcolumntype{Z}{>{\centering\arraybackslash}X}
\label{table}
\centering %
\resizebox{\textwidth}{!}{%
\begin{tabularx}{18.cm}{ZZZZZ}
\toprule[0.1em] %

\multirow{3}*{Acceptance} & \multirow{3}*{Mode} & \multicolumn{3}{c}{\underline{$\;\;\;\;\;\;\;\;\;\;\;\;\;\;\;\;\;\;\;\;$ Integrated cross sections, [nb] $\;\;\;\;\;\;\;\;\;\;\;\;\;\;\;\;\;\;\;\;$ }}  \\[+0.8ex]
                          &                           &  \multirow{2}*{single-diffractive}  & \multirow{2}*{central-diffractive}  & non-diffractive\\[-0.ex]
                          &                           &                                     &                                     & EXP data\\[-0.ex]
                          
\toprule[0.1em]
ATLAS, $|y|< 2.5$           &  \multirow{2}*{$D^{0}+\overline{D^{0}}$} &  \multirow{2}*{3555.22 ($I\!R\!: 25\%$)} & \multirow{2}*{177.35 ($I\!R\!: 43\%$)} & \multirow{2}*{$-$}\\[+0.4ex]
$p_{\perp} > 3.5$ GeV                          & & & & \\[+0.5ex] 
LHCb,$\;$~$2~\!<~\!y~\!<~\!4.5$           &  \multirow{2}*{$D^{0}+\overline{D^{0}}$} &  \multirow{2}*{31442.8 ($I\!R\!: 31\%$)} & \multirow{2}*{2526.7 ($I\!R\!: 50\%$)} & \multirow{2}*{$1488000\pm182000$}\\[+0.4ex]
$p_{\perp} < 8$ GeV                          & & & & \\[-0.ex] 
\hline \\[-1.6ex]
CMS, $|y|< 2.4$           &  \multirow{2}*{$(B^{+}+B^{-})/2$} &  \multirow{2}*{349.18 ($I\!R\!: 24\%$)} & \multirow{2}*{14.24 ($I\!R\!: 42\%$)} & \multirow{2}*{$28100\pm2400\pm2000$}\\[+0.4ex]
$p_{\perp} > 5$ GeV                          & & & & \\[+0.5ex] 
LHCb,$\;$~$2~\!<~\!y~\!<~\!4.5$           &  \multirow{2}*{$B^{+}+B^{-}$} &  \multirow{2}*{867.62 ($I\!R\!: 27\%$)} & \multirow{2}*{31.03 ($I\!R\!: 43\%$)} & \multirow{2}*{$41400\pm1500\pm3100$}\\[+0.4ex]
$p_{\perp} < 40$ GeV                          & & & & \\[-0.ex]  

\hline \\[-1.6ex]
LHCb, $2 < y < 4$           &  \multirow{2}*{$D^{0}\overline{D^{0}}$} &  \multirow{2}*{179.4 ($I\!R\!: 28\%$)} & \multirow{2}*{7.67 ($I\!R\!: 45\%$)} & \multirow{2}*{$6230\pm120\pm230$}\\[+0.4ex]
$3 < p_{\perp} < 12$ GeV                          & & & & \\[-0.ex]

\bottomrule[0.1em]
 
\end{tabularx}
}
}
\end{table}

Numerical predictions of the integrated cross sections for the single- and central-diffractive production of $D^{0}$ and $B^{\pm}$ mesons, including relevant experimental acceptance of the ATLAS, LHCb and CMS detectors,  are collected in Table~\ref{table}. The kinematical cuts are taken to be identical to those which have been already used in the standard non-diffractive measurements. The corresponding experimental cross sections for non-diffractive processes are shown for reference. In the case of inclusive production of single $D$ or $B$ meson the ratio of the diffractive integrated cross sections to the non-diffractive one is about $\sim 2\%$ for single- and only about $\sim 0.07\%$ for central-diffractive mechanism. This ratio is only slightly bigger for $D^{0}\overline{D^{0}}$ pair production, becoming of about $\sim 3\%$ and $0.1\%$, respectively. In addition, the relative contribution of the reggeon-exchange mechanisms to the overall diffractive production cross sections is also shown. This relative contribution is about $\sim 24-31\%$ for single-diffractive ($\frac{I\!R}{I\!P+I\!R}$) and $\sim 42-50\%$ for 
central-diffractive processes ($\frac{I\!PI\!R+I\!RI\!P+I\!RI\!R}{I\!PI\!P+I\!PI\!R+I\!RI\!P+I\!RI\!R}$) for both, charm and bottom flavoured mesons.      

\section{Conclusion}

In the present study we discuss in detail single-
and central- diffractive production of charm and bottom
quark-antiquark pairs as well as open charmed and bottom mesons.
The corresponding cross sections are rather large.
First we have presented cross sections for $c \bar c$ and $b \bar b$
production in single and central production. Several quark-level
differential distributions are shown and discussed. We have compared
pomeron and reggeon contributions.
In order to make predictions which could be compared with future
experimental data we have included hadronization to
charmed ($D$) and bottom ($B$) mesons using hadronization
functions known for other processes.
We have shown several inclusive differential distributions for the
mesons as well as correlations of $D$ and $\bar D$ mesons.
In these calculations we have included detector acceptance
of the ATLAS, CMS and LHCb collaboration experiments.
The production of charmed mesons is interesting
because of the cross section of the order of a few microbarns for ATLAS 
and CMS and of the order of tens of microbarns for the LHCb
acceptance and could be measured.
We have shown that the pomeron contribution is much larger than
the subleading reggeon contribution.

{\bf Acknowledgments}

This study was partially supported by the Polish
National Science Centre grant DEC-2013/09/D/ST2/03724.


\end{document}